\documentstyle[amssymb,11pt]{article}
\input{tcilatex}

\begin{document}

\title{Elementary Excitations in Trapped BEC and Zero Mode Problem }
\author{D.L.Zhou, X.X.Yi, G.R.Jin, Y.H.Su, Y.X.Liu, \& C.P.Sun \\
%EndAName
Institute of Theoretical Physics, Academia Sinica, Beijing, 100080, China}
\date{April, 2000}
\maketitle

\begin{abstract}
We propose a natural expansion of the atomic field operator in studying
elementary excitations in trapped Bose-Einstein Condensation (BEC) system
near T=0K. Based on this expansion, a system of coupled equations for
elementary excitations, which is equivalent to the standard linearized GP
equation, is given to describe the collective excitation of BEC in a natual
way. Applications of the new formalism to the homogeneous case emphasize on
the zero mode and its relevant ground state of BEC.\newline
{\bf PACS number(s):} 03.75.Fi, 05.30.Jp
\end{abstract}

\section{Introduction}

Since the realization of Bose-Einstein Condensation (BEC) in trapped alkali
atomic vapors in 1995\cite{1}, much attentions have been attracted from both
viewpoints of theories and experiments\cite{2}. Among all the recent
developments, the study of collective excitations occupies an important
position in studying the properties of trapped BEC. When the temperature
approaches the critical temperature, the theoretical description of
collective excitation is still proceeding at present\cite{5}. For the case
that the temperature becomes near zero, the linearized GP equations\cite{7,8}
can give correct numerical predictions in agreement with the experiments\cite
{12}. However, the linearized GP equations are difficult to be solved
analytically, and they include confused zero mode problem\cite{3}. In order
to give some new light on this problem, we develop an equivalent formalism
to the linearized GP equations based on a natural expansion of the atomic
field operator.

Our study follows discussions in ref.\cite{3} by Lewenstein and You, and the
main difference is that we adopt specific eigenfunctions to expand the
atomic field. It is noticed that the advantages of our expansion lie in the
following facts: (1) The problem in solving the elementary excitation
becomes the problem in diagonalizing the standard quadratic Hamiltonian of
Boson operators, which has been completely solved in principle before\cite{4}%
. (2) Our formalism adopts a simpler form than those formulated by other
expansion of the atomic field operator. (3) Since only the several lowest
excitations dominate the behavior of trapped BEC system near zero
temperature, the limited-level approximation (which is explained in details
in Sec. 2) is rational in most cases.

This paper is organized as the following: A general formalism for elementary
excitations is present in Sec. 2, and then in Sec. 3 it is applied to the
homogeneous case, which is interest since it can be used to be a good
example to demonstrate the appearance and the physical meaning of the zero
mode clearly. In Sec. 4, a conclusion will be made and some possible
application of our special expansion will be discussed.

\section{General Formalism}

The Hamiltonian of trapped BEC system is 
\begin{equation}
\hat{H}=\int d^3\vec{r}[\hat{\psi}^{\dagger }(\vec{r})(-\frac{\hbar ^2\nabla
^2}{2M}+V_{tr}(\vec{r})-\mu _0)\hat{\psi}(\vec{r})+\frac 12g\hat{\psi}%
^{\dagger }(\vec{r})\hat{\psi}^{\dagger }(\vec{r})\hat{\psi}(\vec{r})\hat{%
\psi}(\vec{r})],  \label{1}
\end{equation}
where $\mu _0$ is the chemical potential, $g=\frac{4\pi \hbar ^2a_s}M$ , $%
a_s $ the s-wave scattering length of the interatomic potential, M the
atomic mass, $V_{tr}(\vec{r})$ the external potential, and $\hat{\psi}(\vec{r%
})$ ($\hat{\psi}^{\dagger }(\vec{r})$) the atomic annihilation (creation)
field operator.

Near $T=0K$, since almost all the atoms occupy the same one-particle quantum
state $\Phi _0(\vec{r})$, it is convenient to separate out the special mode
for the ground state of BEC from the atomic field operator 
\begin{equation}
\hat{\psi}(\vec{r})=\sqrt{N_0}\Phi _0(\vec{r})+\delta \hat{\psi}(\vec{r}).
\label{2}
\end{equation}
where $\delta \hat{\psi}(\vec{r})$ represents the quantum fluctuation of $%
\hat{\psi}(\vec{r})$ relative to$\sqrt{N_0}\Phi _0(\vec{r})$. Notice that,
at present $\int d^3\vec{r}\left\langle \delta \hat{\psi}^{\dagger }(\vec{r}%
)\delta \hat{\psi}(\vec{r})\right\rangle _{ensemble}\ll N_0$, i.e. the
thermal component atoms is negligible. Hence the Hamiltonian can be properly
expanded in $\delta \hat{\psi}(\vec{r})$ series while substituting formula(%
\ref{2}) into formula(\ref{1}). (Only the lower order of $\delta \hat{\psi}(%
\vec{r})$ is important, so up to two orders of $\delta \hat{\psi}(\vec{r})$
is maintained in our consideration.) From the above Hamiltonian, $\Phi _0(%
\vec{r})$ can be determined in zero order of $\delta \hat{\psi}(\vec{r})$ by
making use of variation methods, which satisfy the time-independent GP
equation\cite{6,7} 
\begin{equation}
(-\frac{\hbar ^2\nabla ^2}{2M}+V_{tr}(\vec{r})+gN_0\Phi _0^{*}(\vec{r})\Phi
_0(\vec{r}))\Phi _0(\vec{r})=\mu _0\Phi _0(\vec{r}).  \label{3}
\end{equation}

Combining eqs. (\ref{1}-\ref{3}), and ignoring the terms of the third and
fourth order of $\delta \hat{\psi}(\vec{r})$ in eq.(\ref{1}), the
Hamiltonian can be simplified as, 
\begin{eqnarray}
\hat{H} &\simeq &-\frac 12gN_0^2\int d^3\vec{r}\Phi _0^{*}(\vec{r})\Phi
_0^{*}(\vec{r})\Phi _0(\vec{r})\Phi _0(\vec{r})  \nonumber \\
&&+\int d^3\vec{r}\delta \hat{\psi}^{\dagger }(\vec{r})(-\frac{\hbar
^2\nabla ^2}{2M}+V_{tr}(\vec{r})-\mu _0+2gN_0\Phi _0^{*}(\vec{r})\Phi _0(%
\vec{r}))\delta \hat{\psi}(\vec{r})  \nonumber \\
&&+\frac 12gN_0\int d^3\vec{r}\Phi _0^{*}(\vec{r})\Phi _0^{*}(\vec{r})\delta 
\hat{\psi}(\vec{r})\delta \hat{\psi}(\vec{r})  \nonumber \\
&&+\frac 12gN_0\int d^3\vec{r}\delta \hat{\psi}^{\dagger }(\vec{r})\delta 
\hat{\psi}^{\dagger }(\vec{r})\Phi _0(\vec{r})\Phi _0(\vec{r}).  \label{4}
\end{eqnarray}

In order to further simplify the Hamiltonian, we select a set of special
complete wave functions $\left\{ \Phi _n(\vec{r})\right\} $ to expand the
atomic field operator, 
\begin{equation}
\hat{\psi}(\vec{r})=\sum_{n\neq 0}\hat{a}_n\Phi _n(\vec{r})+\hat{A}_0\Phi _0(%
\vec{r}),  \label{5}
\end{equation}
where$\left\{ \Phi _n(\vec{r})\right\} $ satisfy the following equations\cite
{11} 
\begin{equation}
(-\frac{\hbar ^2\nabla ^2}{2M}+V_{tr}(\vec{r})+gN_0\Phi _0^{*}(\vec{r})\Phi
_0(\vec{r}))\Phi _n(\vec{r})=\mu _n\Phi _n(\vec{r}),  \label{6}
\end{equation}
$\hat{a}_n$ is the annihilation boson operator of the single state $\Phi _n(%
\vec{r})$ $(n\neq 0)$, $\hat{A}_0$ the annihilation boson operator of the
single state $\Phi _0(\vec{r})$. Due to the fact that $(-\frac{\hbar
^2\nabla ^2}{2M}+V_{tr}(\vec{r})+gN_0\Phi _0^{*}(\vec{r})\Phi _0(\vec{r}))$
is Hermitian, its eigenstates $\Phi _n(\vec{r})$ $(n=0,1,\cdots )$ are
orthorgonal and complete 
\begin{equation}
\int d^3\vec{r}\Phi _m^{*}(\vec{r})\Phi _n(\vec{r})=\delta _{mn}.  \label{7}
\end{equation}
In most cases, only the lower part of elementary excitations determines the
properties of the system near zero temperature, so it is a good
approximation to hold only the lower $f$-level of eq.(\ref{6}), i.e. $%
n=\left\{ 0,1,\cdots ,f-1\right\} $. In addition, when $f\rightarrow \infty $%
, the present treatment becomes rigorous in principle.

Substituting eq. (\ref{5}) into eq. (\ref{2}), we obtain 
\begin{equation}
\delta \hat{\psi}(\vec{r})=\sum_{n=0}^{f-1}\hat{a}_n\Phi _n(\vec{r}),
\label{8}
\end{equation}
where 
\begin{equation}
\hat{a}_0=\hat{A}_0-\sqrt{N_0}.  \label{9}
\end{equation}

In terms of eqs. (\ref{8},\ref{6}), the Hamiltonian (\ref{4}) can be
reexpressed as 
\begin{eqnarray}
\hat{H} &\simeq &-\frac{B_{00}}2N_0+\sum_{m,n}A_{mn}\hat{a}_m^{\dagger }\hat{%
a}_n  \nonumber \\
&&+\frac 12\sum_{m,n}(B_{mn}^{*}\hat{a}_m\hat{a}_n+B_{mn}\hat{a}_m^{\dagger }%
\hat{a}_n^{\dagger }),  \label{10}
\end{eqnarray}
where 
\begin{equation}
A_{mn}=(\mu _m-\mu _0)\delta _{mn}+d_{mn},  \label{11}
\end{equation}
\begin{eqnarray}
B_{mn} &=&gN_0\int d^3\vec{r}\Phi _0(\vec{r})\Phi _0(\vec{r})\Phi _m^{*}(%
\vec{r})\Phi _n^{*}(\vec{r}),  \label{12} \\
d_{mn} &=&gN_0\int d^3\vec{r}\Phi _m^{*}(\vec{r})\Phi _0^{*}(\vec{r})\Phi _0(%
\vec{r})\Phi _n(\vec{r}).  \label{13}
\end{eqnarray}

The scheme of diagonalizing the Hamiltonian as the form (\ref{10}) has been
extensively studied by Blaizot and Ripka\cite{4}. Here, their results will
be briefly reviewed in the following for the use in our present discussion.

First, one write the Hamiltonian in a compact form 
\begin{equation}
\hat{H}\simeq -\frac{B_{00}}2N_0+\frac 12\alpha ^{\dagger }M\alpha -\frac
12trA  \label{14}
\end{equation}
by introducing the vector operator 
\begin{equation}
\alpha ^{\dagger }=\left( 
\begin{array}{ll}
\hat{a}^{\dagger } & \hat{a}
\end{array}
\right) =\left( 
\begin{array}{llllllll}
\hat{a}_0^{\dagger } & \hat{a}_1^{\dagger } & \cdots  & \hat{a}%
_{f-1}^{\dagger } & \hat{a}_0 & \hat{a}_1 & \cdots  & \hat{a}_{f-1}
\end{array}
\right) ,  \label{15}
\end{equation}
the $2f\times 2f$ coefficients matrix 
\begin{equation}
M=\left( 
\begin{array}{ll}
A & B \\ 
B^{*} & A^{*}
\end{array}
\right) ,A=A^{\dagger },B=\tilde{B}.  \label{16}
\end{equation}
In ref.\cite{4}, it is assumed that the matrix $M$ is semi-definite
positive. However, in our discussion, this condition on the matrix $M$ is
not necessary since it is only a condition for the stability of the mode of
the ground state of BEC\cite{9}.

Second, a unitary canonical transformation is carried out to diagonalize the
Hamiltonian, 
\begin{equation}
\beta =T\alpha  \label{18}
\end{equation}
where the operator vector 
\begin{equation}
\beta ^{\dagger }=\left( 
\begin{array}{ll}
\hat{b}^{\dagger } & \hat{b}
\end{array}
\right) =\left( 
\begin{array}{llllllll}
\hat{b}_0^{\dagger } & \hat{b}_1^{\dagger } & \cdots & \hat{b}%
_{f-1}^{\dagger } & \hat{b}_0 & \hat{b}_1 & \cdots & \hat{b}_{f-1}
\end{array}
\right) ,  \label{19}
\end{equation}
the transformation matrix and its inverse matrix 
\begin{equation}
T=\left( 
\begin{array}{ll}
X^{*} & -Y^{*} \\ 
-Y & X
\end{array}
\right) ,T^{-1}=\left( 
\begin{array}{ll}
\tilde{X} & Y^{\dagger } \\ 
\tilde{Y} & X^{\dagger }
\end{array}
\right) ,  \label{20}
\end{equation}
the matrix satisfy the canoical condition 
\begin{equation}
T\eta T^{\dagger }\eta =1,  \label{21}
\end{equation}
and the matrix 
\begin{equation}
\eta =\left( 
\begin{array}{ll}
1_f & 0 \\ 
0 & -1_f
\end{array}
\right) .  \label{22}
\end{equation}

If the elements of the $f\times f$ matrixes $X$ and $Y$ satisfy the
following conditions\cite{4,14} 
\begin{equation}
\eta MV^n=\varpi _nV^n,  \label{23}
\end{equation}
where 
\[
V^n=\left( 
\begin{array}{l}
X^n \\ 
Y^n
\end{array}
\right) ,X_i^n=X_{ni},Y_i^n=Y_{ni},(i=1,2,\cdots ,f) 
\]
especially, if the above equation have one special solution $\left\{ \varpi
_0=0,V^0=P\right\} $, i.e. 
\begin{equation}
\eta MP=0  \label{24}
\end{equation}
the Hamiltonian can be written as 
\begin{equation}
\hat{H}\simeq -\frac{B_{00}}2N_0+\sum_{n=1}^{f-1}\omega _n\hat{b}_n^{\dagger
}\hat{b}_n+\frac{\wp ^2}{2\mu }+\frac 12\sum_n\omega _n-\frac 12trA
\label{25}
\end{equation}
where 
\begin{equation}
\wp \equiv \alpha ^{\dagger }\eta P,  \label{26}
\end{equation}
and $\mu $ is a positive constant, which can be determined by the following
conditions 
\begin{eqnarray}
\eta MQ &=&-i\frac P\mu ,  \label{a1} \\
Q^{\dagger }\eta P &=&i,  \label{a2}
\end{eqnarray}
where the vector $Q$ is orthorgonal to all the eigenvectors of the matrix $%
\eta M$.

In Hamiltonian (\ref{25}), the second term is the Hamiltonian of a system of
independent oscillators, which represent elementary excitations of the
system; However, the third term has the form of a free kinetic energy, which
is connected with a collective motion in Fock space arising from a broken $%
U(1)$ symmetry in the procedure of the mean field approximation\cite{4}.
Usually, the third term is termed with {\it spurious state}\cite{4}{\it \ }%
or zero mode\cite{3} for the corresponding eigenvalue and the norm of the
vector $P$ both are zero. The physical meaning of this term will be
discussed in details in Sec. 3.

In fact, we can determine the vector $P$ through observation. Notice that in
principle, we can solve the the eigenvalues $\left\{ \omega _n\right\} $ and
the corresponding eigenvectors defined by $\left\{ X_m^n,Y_m^n\right\} $ of
eqs. (\ref{23}), which consist of $2\times f\times f$ homogeneous linear
equations. Obviously, the eigenvalue $\omega _0=0$ and the corresponding
vector $P$ denoted by $\left\{ X_m^0=\delta _{m0},Y_m^0=-\delta
_{m0}\right\} $ is a specific solution of the above equations. Hence 
\begin{equation}
\wp =\hat{a}_0+\hat{a}_0^{\dagger },  \label{27}
\end{equation}
which is in agreement with that in ref.\cite{3}.

From eqs. (\ref{25},\ref{27}), a direct conclusion is that the approximate
vacuum state $\left| Vac\right\rangle =\prod_i\otimes \left|
Vac\right\rangle _i$ of BEC satisfies the following conditions 
\begin{eqnarray}
\hat{b}_i\left| Vac\right\rangle _i &=&0,  \label{28} \\
\wp \left| Vac\right\rangle _0 &=&0.  \label{29}
\end{eqnarray}

To sum up, in this section, we give a new formalism for elementary
excitations in trapped BEC, which is equivalent to the standard linearized
GP equation. This equivalence can be easily verified when the complex wave
functions $\left( 
\begin{array}{l}
u(\vec{r}) \\ 
v(\vec{r})
\end{array}
\right) $ in traditional method are expanded with the specific complete wave
functions defined by eqs.(\ref{6}). In this sense, our formalism is a
specific representation of the traditional method.

\section{Homogeneous case $V_{tr}(\vec{r})=0$}

In this section, the general formalism obtained in the above section will be
demonstrated in the homogeneous case.

In this case, the ground wave function satisfy 
\begin{equation}
(-\frac{\hbar ^2\nabla ^2}{2M}+gN_0\Phi _0^{*}(\vec{r})\Phi _0(\vec{r}))\Phi
_0(\vec{r})=\mu _0\Phi _0(\vec{r}).
\end{equation}
Therefore, the ground wave function and the chemical potential are given by 
\begin{equation}
\Phi _0(\vec{r})=\frac 1{\sqrt{V}},\mu _0=g\frac{N_0}V.
\end{equation}
Eq. (\ref{6}) which give the complete wave functions now becomes 
\begin{equation}
(-\frac{\hbar ^2\nabla ^2}{2M}+\mu _0)\Phi _{\vec{k}}(\vec{r})=\mu _{_{\vec{k%
}}}\Phi _{\vec{k}}(\vec{r})
\end{equation}
By solving the above equations, the eigen wave functions and the
corresponding eigen values are given by 
\begin{eqnarray}
\Phi _{_{\vec{k}}}(\vec{r}) &=&\frac 1{\sqrt{V}}e^{i\vec{k}\cdot \vec{r}}, 
\nonumber \\
\mu _{_{\vec{k}}} &=&\mu _0+\frac{\hbar ^2k^2}{2M}.
\end{eqnarray}
Formulas (\ref{12},\ref{13}) become 
\begin{eqnarray}
B_{\vec{k}\vec{k}^{\prime }} &=&gN_0\int d^3\vec{r}\Phi _0(\vec{r})\Phi _0(%
\vec{r})\Phi _{\vec{k}}^{*}(\vec{r})\Phi _{\vec{k}^{\prime }}^{*}(\vec{r})=g%
\frac{N_0}V\delta _{\vec{k},-\vec{k}^{\prime }},  \nonumber \\
d_{\vec{k}\vec{k}^{\prime }} &=&gN_0\int d^3\vec{r}\Phi _{\vec{k}}^{*}(\vec{r%
})\Phi _0^{*}(\vec{r})\Phi _0(\vec{r})\Phi _{\vec{k}^{\prime }}(\vec{r})=g%
\frac{N_0}V\delta _{\vec{k},\vec{k}^{\prime }}.
\end{eqnarray}
Due to the fact the mode denoted by $\vec{k}$ is only coupled to the mode
denoted by $-\vec{k}$ implied by the above equation, eqs.(\ref{23}) can be
simplified as

\begin{eqnarray}
(\frac{\hbar ^2k^2}{2M}+g\frac{N_0}V-\omega _{\vec{k}^{\prime }})X_{\vec{k}%
}^{\vec{k}^{\prime }}+g\frac{N_0}VY_{-\vec{k}}^{\vec{k}^{\prime }} &=&0 
\nonumber \\
(\frac{\hbar ^2k^2}{2M}+g\frac{N_0}V+\omega _{\vec{k}^{\prime }})Y_{\vec{k}%
}^{\vec{k}^{\prime }}+g\frac{N_0}VX_{-\vec{k}}^{\vec{k}^{\prime }} &=&0
\end{eqnarray}

When $\vec{k}^{\prime }\neq 0$, the eigenvalue can be calculated by
requiring that the above equations have nontrivial solution, 
\begin{equation}
\omega _{\vec{k}^{\prime }}=\omega _{-\vec{k}^{\prime }}=\sqrt{(\frac{\hbar
^2k^{^{\prime }2}}{2M}+g\frac{N_0}V)^2-(g\frac{N_0}V)^2}
\end{equation}
The corresponding annihilation operators of the elementary excitation are 
\[
\hat{b}_{\vec{k}^{\prime }}=\sqrt{\frac 12(\frac{\frac{\hbar ^2k^{^{\prime
}2}}{2M}+g\frac{N_0}V}{\omega _{\vec{k}^{\prime }}}+1)}\hat{a}_{\vec{k}%
^{\prime }}-\sqrt{\frac 12(\frac{\frac{\hbar ^2k^{^{\prime }2}}{2M}+g\frac{%
N_0}V}{\omega _{\vec{k}^{\prime }}}-1)}\hat{a}_{-\vec{k}^{\prime }}^{\dagger
}
\]
or 
\[
\hat{b}_{-\vec{k}^{\prime }}=\sqrt{\frac 12(\frac{\frac{\hbar ^2k^{^{\prime
}2}}{2M}+g\frac{N_0}V}{\omega _{\vec{k}^{\prime }}}+1)}\hat{a}_{-\vec{k}%
^{\prime }}-\sqrt{\frac 12(\frac{\frac{\hbar ^2k^{^{\prime }2}}{2M}+g\frac{%
N_0}V}{\omega _{\vec{k}^{\prime }}}-1)}\hat{a}_{\vec{k}^{\prime }}^{\dagger }
\]
Clearly, it comes back to the familiar form\cite{2}, which supports that our
formalism is equivalent to the traditional one.

Since the operators $\hat{a}_0$ and $\hat{a}_0^{\dagger }$ are only coupled
each other, we can limit ourself in the subspace of the wave vector $\vec{k}%
^{\prime }=0$. The matrix $\eta M$ in this subspace is

\[
\eta M=\left( 
\begin{array}{ll}
g\frac{N_0}V & g\frac{N_0}V \\ 
-g\frac{N_0}V & -g\frac{N_0}V
\end{array}
\right) .
\]
Obviously, the eigen vector $P$ of the zero mode and the corresponding
momentum operator $\wp $ are obtained as 
\[
P=\left( 
\begin{array}{l}
1 \\ 
-1
\end{array}
\right) ,\wp =\hat{a}_0+\hat{a}_0^{\dagger }.=(\hat{A}_0+\hat{A}_0^{\dagger
})-2\sqrt{N_0}.
\]
According to eqs. (\ref{a1},\ref{a2}), the other independent vector $Q$ and
the constant $\mu $ are obtained as 
\[
Q=\frac{-i}2\left( 
\begin{array}{l}
1 \\ 
1
\end{array}
\right) ,\mu ^{-1}=g\frac{N_0}V.
\]
In sum, the Hamiltonian can be written as 
\begin{equation}
\hat{H}\simeq -\frac{gN_0^2}{2V}+\sum_{\vec{k}\neq 0}\omega _{\vec{k}}\hat{b}%
_{\vec{k}}^{\dagger }\hat{b}_{\vec{k}}+\frac{\wp ^2}{2\mu }+\frac 12\sum_{%
\vec{k}\neq 0}\omega _{\vec{k}}-\frac 12\sum_{\vec{k}}(\frac{\hbar ^2k^2}{2M}%
+g\frac{N_0}V).  \label{a4}
\end{equation}

Now, the approximate vacuum state $\left| Vac\right\rangle =\prod_{\vec{k}%
}\otimes \left| Vac\right\rangle _{\vec{k}}$ of BEC can be obtained
analytically by solving eqs.(\ref{28},\ref{29}). When the wave vector $\vec{k%
}\neq 0$, the vacuum state $\left| Vac\right\rangle _{\vec{k}}$ is given by
solving the equations $\hat{b}_{\vec{k}}\left| Vac\right\rangle _{\vec{k}}=%
\hat{b}_{-\vec{k}}\left| Vac\right\rangle _{\vec{k}}=0$, 
\[
\left| Vac\right\rangle _{\vec{k}}=\sum_nA_{\vec{k}}\left( \frac{\frac{\hbar
^2k^2}{2M}+g\frac{N_0}V-\omega _{\vec{k}}}{\frac{\hbar ^2k^2}{2M}+g\frac{N_0}%
V+\omega _{\vec{k}}}\right) ^{\frac n2}\left| n\right\rangle _{\vec{k}%
}\otimes \left| n\right\rangle _{-\vec{k}},
\]
where $A_{\vec{k}}$ is the constant of normalization, the vector $\left|
n\right\rangle _{\vec{k}}$ ($\left| n\right\rangle _{-\vec{k}}$) is the
eigenstate of the number operator $\hat{a}_{\vec{k}}^{\dagger }\hat{a}_{\vec{%
k}}$ ($\hat{a}_{-\vec{k}}^{\dagger }\hat{a}_{-\vec{k}}$). When the wave
vector $\vec{k}=0$, $\wp \left| Vac\right\rangle _0=0$. Therefore,

\begin{equation}
\left| Vac\right\rangle _0=\frac 1{\sqrt{2\pi }}\sum_n\left| n\right\rangle
_0\int dxe^{i\sqrt{2N_0}x}\langle x|n\rangle ^{*},
\end{equation}
where the state vector $\left| n\right\rangle _0$ is the eigen state of the
number operator $\hat{A}_0^{\dagger }\hat{A}_0$, 
\[
\langle x|n\rangle =[\frac 1{\sqrt{\pi }2^nn!}]^{\frac 12}e^{-\frac
12x^2}H_n(x), 
\]
and $H_n(x)$ is Hermit polynomial.

In this section, we explicitly solve the elementary excitations of BEC in
the homogeneous case. In this case, it is easy to see that the term of the
zero mode in Hamiltonian (\ref{a4}) originate from the quantum fluctuation
of the mode denoted by macroscopic wave function $\Phi _0(\vec{r})$. In
fact, if we adopt the usual Bogoliubov approximation, i.e. $\hat{A}%
_0\backsim \hat{A}_0^{\dagger }\backsim \sqrt{N_0}$, thus the momentum
operator $\wp \equiv 0$. However, we have no specific reasons to ignore this
quantum fluctuation while maintaining those of the other modes. In our
treatment, the conservation of the particle number is destroyed, which is
easily seen from the approximate vacuum state $\left| Vac\right\rangle $.
The kinetic term appearing in the Hamiltonian is originated from this
symmetry breaking, which represents a collective motion, not an intrinsic
elementary excitation of the system\cite{4}.

\section{Conclusion}

In this paper, based on a natural choice of the complete wave functions, we
expand the atomic field operator and obtain a new formalism for the
excitations of trapped BEC system near zero temperature. We argue that our
formalism is equivalent to the standard linearized GP equation. In terms of
this formalism, we illustrate the relation between the zero mode and the
other excited modes. Essentially, the zero mode originates from the quantum
fluctuations of the mode denoted by the condensate wave function. When
applicating the formalism to the homogeneous case, the formalism comes back
to the usual Bogoliubov excitation spectrum, which identifies our theory..
Especially, in this case, the physical meaning of zero mode become obvious
and the ground state of BEC can be calculate explicitly up to second order
of the quantum fluctuations.\newline
{\bf ACKNOWLEDGMENT: This work is supported by NSF of China.} 


\medskip 

\begin{thebibliography}{99}
\bibitem{1}  M.H. Anderson et al., Science {\bf 269, }198 (1995); C.C.
Bradley et al., Phys. Rev. Lett. {\bf 75}, 1687 (1995); K.B. Davis et al.,
Phys. Rev. Lett. {\bf 75}, 3969 (1995).

\bibitem{2}  F. Dalfovo et al, Rev. Mod. Phys. {\bf 71, }463 (1999).

\bibitem{3}  M. Lewenstein and L. You, Phys. Rev. Lett. {\bf 77}, 3489
(1997).

\bibitem{4}  J.-P. Blaizot and G. Ripka, {\it Quantum Theory of Finite
Systems} (MIT Press, Cambridge, MA, 1986).

\bibitem{5}  M. Rusch and K. Burnett, Phys. Rev. A {\bf 59}, 3851 (1999).

\bibitem{6}  V.L. Ginzburg and L.P. Pitaevskii, Sov. Phys. JETP {\bf 7}, 858
(1958); E.P. Gross, J. Math. Phys. (N.Y.) {\bf 4}, 195 (1963).

\bibitem{7}  L.P. Pitaevskii, Sov. Phys. JETP {\bf 13}, 451 (1961).

\bibitem{8}  A.L. Fetter, Ann. Phys. (N. Y.) {\bf 70}, 67 (1972);
A.L.Fetter, cond-mat/9811366.

\bibitem{9}  C.K. Law et al. Phys. Rev. Lett. {\bf 79}, 3105 (1997). In this
paper, they define that when the matrix M is not semi-definite positive, the
system will become unstable.

\bibitem{10}  H. Shi and A. Griffin, Phys. Rep. {\bf 304}, 1 (1998).

\bibitem{11}  J. Williams et al., Phys. Rev. A {\bf 61}, 033612(2000). In
the article, we found a similar expansion, however, it has essential
difference, mainly lying in the fact that in their article the macroscopic
wave function is time dependent.

\bibitem{12}  M. Edward et al., Phys. Rev. Lett. {\bf 77}, 1671 (1997).

\bibitem{13}  A. Imamo\={g}lu et al., Phys. Rev. Lett. {\bf 78}, 2511 (1997).

\bibitem{14}  Note: Although eq.(\ref{23}) has $2f$ independent solutions,
however, if the canonical condition eq. (\ref{21}) is considered, no more
than $f$ independent solutions exist.
\end{thebibliography}
\end{document}